\documentclass{article}
\usepackage[a4paper, total={6in, 10in}]{geometry}
\usepackage{graphicx} 
\usepackage{multirow}
\usepackage{siunitx}
\usepackage{url}
\usepackage{hyperref}
\usepackage{authblk}
\usepackage{subcaption}
\makeatletter
\def\bstctlcite{\@ifnextchar[{\@bstctlcite}{\@bstctlcite[@auxout]}}
\def\@bstctlcite[#1]#2{\@bsphack
  \@for\@citeb:=#2\do{%
    \edef\@citeb{\expandafter\@firstofone\@citeb}%
    \if@filesw\immediate\write\csname #1\endcsname{\string\citation{\@citeb}}\fi}%
  \@esphack}
\makeatother

\title{Multi-Tiered Self-Contrastive Learning for Medical Microwave Radiometry (MWR) Breast Cancer Detection}
\author[1]{Christoforos Galazis}
\author[2]{Huiyi Wu}
\author[3,4,5,*]{Igor Goryanin}
\affil[1]{Department of Computing, Imperial College London, London, United Kingdom}
\affil[2]{National Heart \& Lung Institute, Imperial College London, London, United Kingdom}
\affil[3]{School of Informatics, University of Edinburgh, Edinburgh, United Kingdom}
\affil[4]{Okinawa Institute Science and Technology, Okinawa, Japan}
\affil[5]{MMWR Ltd, United Kingdom}
\affil[*]{igor.goryanin@ed.ac.uk}

\date{}

\providecommand{\keywords}[1]{\textbf{\textit{Keywords }} #1}

\begin{document}
\bstctlcite{IEEEexample:BSTcontrol}

\maketitle

\begin{abstract}
Improving breast cancer detection and monitoring techniques is a critical objective in healthcare, driving the need for innovative imaging technologies and diagnostic approaches. This study introduces a novel multi-tiered self-contrastive model tailored for microwave radiometry (MWR) in breast cancer detection. Our approach incorporates three distinct models: Local-MWR (L-MWR), Regional-MWR (R-MWR), and Global-MWR (G-MWR), designed to analyze varying sub-regional comparisons within the breasts. These models are integrated through the Joint-MWR (J-MWR) network, which leverages self-contrastive results at each analytical level to improve diagnostic accuracy. Utilizing a dataset of 4,932 female patients, our research demonstrates the efficacy of our proposed models. Notably, the J-MWR model achieves a Matthew's correlation coefficient of 0.74 ± 0.018, surpassing existing MWR neural networks and contrastive methods. These findings highlight the potential of self-contrastive learning techniques in improving the diagnostic accuracy and generalizability for MWR-based breast cancer detection. This advancement holds considerable promise for future investigations into enabling point-of-care testing. The source code is available at: \url{https://github.com/cgalaz01/self_contrastive_mwr}.
\end{abstract}


\section{Introduction}

\begin{figure*}[!hbtp]
\centering
  \includegraphics[width=1.0\textwidth]{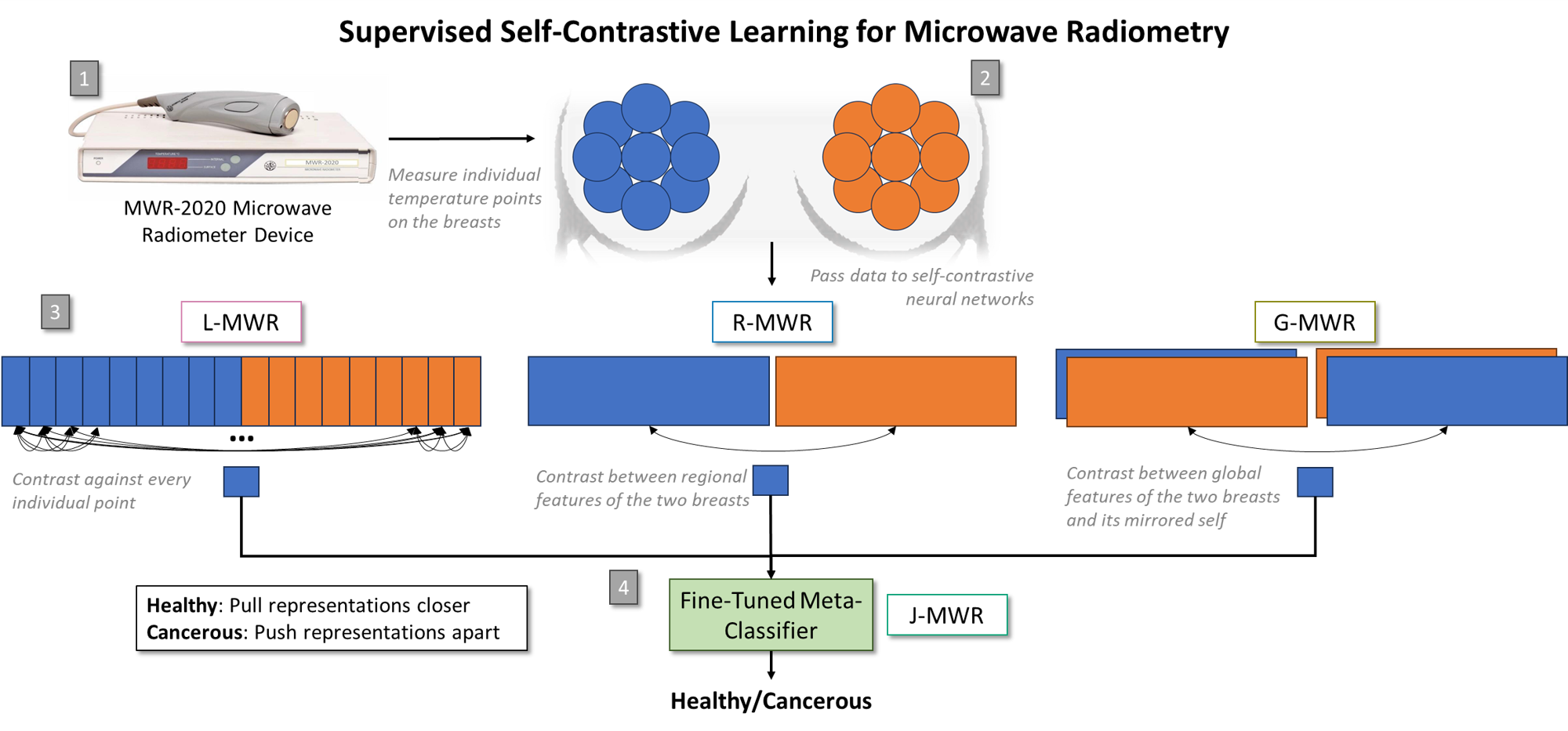}
  \caption{Overview of the proposed self-contrastive models for breast cancer detection. 1) A microwave radiometry (MWR) device, MWR-2020, is used to capture skin and internal temperatures at 2) predetermined locations on the breasts. 3) The data is processed through three hierarchical levels of supervised self-contrastive models: Local-MWR (L-MWR) compares individual temperature points, Regional-MWR (R-MWR) compares temperature features between the left and right breasts, and Global-MWR (G-MWR) compares features of both breasts with their inverted counterparts. 4) The outputs from these pre-trained models are aggregated and fine-tuned using a meta-classifier, Joint-MWR (J-MWR).}
  \label{fig:summary}
\end{figure*}

Breast cancer, characterized by the uncontrolled and rapid growth of cells due to genetic mutations, significantly impacts global health, with one of the highest incidence rates among cancers. In 2020 alone, breast cancer accounted for an estimated 2.3 million new cases and nearly 700,000 deaths, making it the leading cause of cancer-related mortality among women \cite{sung2021global}. Alarmingly, future projections suggest a continued rise in both the incidence and mortality rates associated with breast cancer \cite{arnold2022current}.

Early detection plays a pivotal role in reducing mortality rates and mitigating the healthcare burden. In this context, microwave imaging radiometry (MWR) emerges as a promising imaging modality that passively captures the natural microwave emissions from human tissues \cite{goryanin2020passive}. MWR’s applications span diverse clinical areas, including breast imaging \cite{goryanin2020passive,vesnin2017modern,fisher2022passive,khoperskov2022improving,wang2023microwave,bhargava2024microwave}, brain temperature monitoring \cite{shevelev2022using,shevelev2023correction,shevelev2023diagnostics,hossain2023lightweight}, lung diagnostics \cite{osmonov2021passive,emilov2023diagnostic}, vein assessment \cite{levshinskii2022using}, and musculoskeletal evaluations \cite{tarakanov2022passive,tarakanov2023age}.

In cancer diagnostics, MWR capitalizes on the increased metabolic rate of cancerous tissues, which emit higher levels of heat compared to normal tissues \cite{vesnin2017modern}. Compared to traditional screening methods, MWR offers distinct advantages, including its non-invasive, safe, mobile, and cost-effective nature \cite{goryanin2020passive}. However, the novelty of MWR in breast cancer diagnostics presents challenges, particularly in data interpretation and integrating this technology into established medical workflows. Addressing these challenges necessitates the development of artificial intelligence models to optimize and streamline MWR's clinical applications.



In recent years, significant advancements have been made in using machine learning and deep learning techniques to improve the diagnostic accuracy of MWR-based breast cancer detection. Studies have demonstrated the efficacy of traditional machine learning methods such as support vector machines (SVM), random forests, and Bayesian classifiers \cite{levshinskii2020application,galazis2019application,losev2022some}. Neural network-based approaches have also shown promise \cite{levshinskii2020application,galazis2019application,losev2021neural,losev2022artificial}. For instance, learnable architectures tailored for lightweight, resource-efficient neural networks \cite{li2022dynamic} and the integration of Kohonen’s self-organizing maps with machine learning algorithms \cite{khoperskov2022improving} have advanced the field. Fuzzy analysis techniques have further contributed to improving system efficacy \cite{germashev2021fuzzy,germashev2023hierarchical}.

Prior to the adoption of data-driven methods, hand-engineered features were proposed based on expert knowledge to identify temperature asymmetries within the breast \cite{zenovich2016algoritmy,losev2017intellektualnyy,fisher2022passive}. These features, aimed at analyzing mammary gland thermography, can be categorized into five groups: 1) temperature asymmetry between the two glands; 2) increased temperature dispersion within a single gland; 3) detection of abnormally high temperatures in the nipple relative to other gland areas; 4) relationships between surface and depth temperature measurements; and 5) features derived from comparing the two glands.

Despite these advancements, integrating domain knowledge into data-driven methodologies remains a significant challenge. Combining hand-engineered features with machine learning models is crucial for improving both diagnostic performance and generalizability, which are essential for ensuring the clinical applicability of MWR in diverse and evolving scenarios.

This paper introduces a novel supervised neural network, the Joint-MWR (J-MWR), which incorporates three hierarchical self-contrastive learning networks to improve MWR-based breast cancer detection. As illustrated in Figure \ref{fig:summary}, J-MWR employs self-contrastive learning to analyze and contrast different regions within the same instance, diverging from traditional methods that compare data across cases \cite{zenovich2016algoritmy,losev2017intellektualnyy,fisher2022passive}. By embedding domain knowledge into a data-driven framework, J-MWR bridges the gap between traditional feature extraction and the scalability of neural networks. Experimental results demonstrate that J-MWR outperforms state-of-the-art MWR models and widely used batch-wise contrastive learning techniques, underscoring its potential as a robust solution for breast cancer detection.

\section{Materials and Methods}
\subsection{Data}
The dataset used in this study was collected using the MWR-2020 dual-band point-of-care device, developed by MMWR Ltd\footnote{\url{https://www.mmwr.co.uk/}}. This device monitors both infrared (skin surface temperature) and microwave (internal tissue temperature) emissions. Operating within a microwave frequency range of 3.5 to 4.2 GHz, it is capable of penetrating to tissue depths of 3 to 7 cm. Temperature measurements are acquired with an accuracy of $\pm {0.2}^\circ C$.

\begin{figure*}[!hbtp]
\centering
  \includegraphics[width=0.8\textwidth]{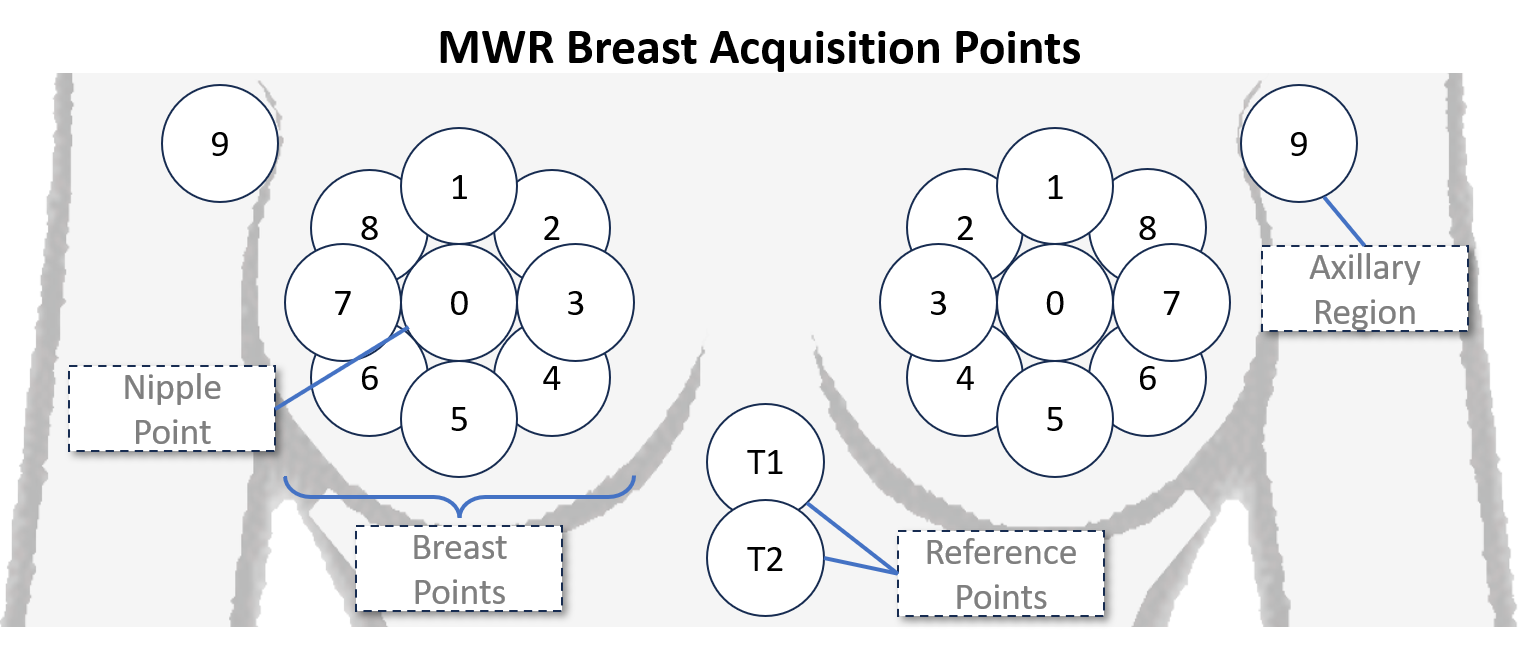}
  \caption{An illustration of the skin and internal acquisition points on the breasts. Point 0 represents the nipple, points 1–8 are arranged equidistantly around the nipple, point 9 corresponds to the axillary region, and reference points T1 and T2 are located beneath the chest.}
  \label{fig:acquisition_points}
\end{figure*}

Data collection involved recording temperatures at 22 specific locations for each case, as depicted in Figure \ref{fig:acquisition_points}. For each breast, eight points were measured equidistantly around the nipple, alongside one measurement directly at the nipple. Additional measurements included the left and right axillary reference regions and two reference points beneath the chest. This resulted in a total of 44 temperature readings (22 skin temperatures and 22 internal temperatures) per case, which served as the input features for the models.

\begin{figure*}[!hbtp]
\centering
  \includegraphics[width=1.0\textwidth]{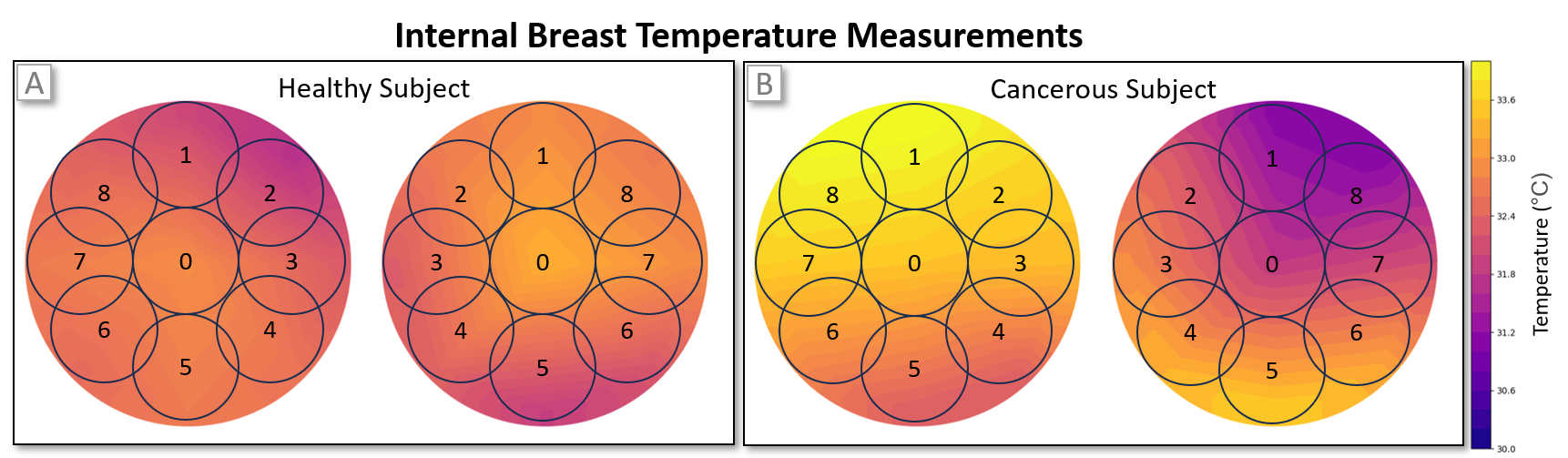}
  \caption{Comparison of internal breast temperature profiles between a healthy case (Panel \textbf{A}) and a high-growth-rate cancerous case (Panel \textbf{B}). In the healthy case, temperature profiles exhibit no significant asymmetries. On the other hand, in the cancerous case, elevated temperatures are observed in regions 1 and 8 of the right gland, indicating localized abnormalities.}
  \label{fig:temperature_example}
\end{figure*}

The dataset comprises measurements from 4,932 female cases, collected across multiple clinical centers with full ethical approval and informed consent. The data was anonymized prior to analysis. Of these cases, 4,384 were classified as healthy (label = 0) by clinical experts, while 548 were labeled as cancerous (label = 1). Examples of a healthy and a cancerous case are shown in Figure \ref{fig:temperature_example}.A and \ref{fig:temperature_example}.B, respectively. The dataset was randomly divided into training (60\%), validation (20\%), and testing (20\%) subsets, maintaining class balance. Model evaluation was conducted on the testing set after completing model development and experimentation phases.

\subsection{Models}
The models in this study share consistent initialization and training settings to ensure comparability. Weights were initialized using the Glorot uniform method \cite{glorot2010understanding}, and biases were set to zero. The Adam optimizer \cite{kingma2014adam} was used for weight optimization, with a starting learning rate of 0.0001 and parameters $\beta_1 = 0.9$ and $\beta_2 = 0.999$. These settings were chosen based on findings from previous studies \cite{levshinskii2020application,galazis2019application} using similar data. To further improve training stability, the learning rate was reduced by a factor of 0.1 if the validation loss did not improve after five consecutive epochs. The batch size was set to 4, determined experimentally as optimal across all models.

To address the class imbalance between healthy and cancerous cases, a class-balanced binary cross-entropy loss function was used. This approach ensured that the models maintained equal focus on both majority (healthy) and minority (cancerous) classes during training.

\subsection{Base Model}
The base model was derived from a neural network architecture previously proposed for MWR analysis \cite{levshinskii2020application}. Modifications were made to accommodate the larger dataset used in this study, optimizing the model for improved performance.

\begin{figure*}[!ht]
\centering
  \includegraphics[width=1.0\textwidth]{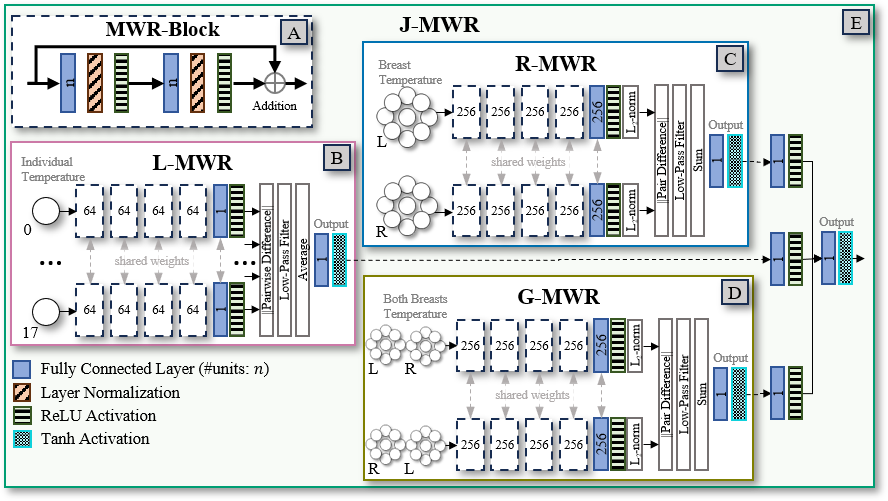}
  \caption{Overview of the proposed multi-tiered self-contrastive MWR models for breast cancer detection. \textbf{A)} The shared MWR-Block, a common residual block used across all networks. \textbf{B)} L-MWR network, which performs point-to-point comparisons within each breast. \textbf{C)} R-MWR network, which conducts comparisons between the left and right breasts. \textbf{D)} G-MWR network, which compares each breast with its positional inverse counterpart. \textbf{E)} J-MWR network, which integrates predictions from L-MWR, R-MWR, and G-MWR to produce the final prediction.}
  \label{fig:models}
\end{figure*}

The base model consists of four residual fully connected (FC) modules, referred to as MWR-Blocks, as shown in Figure \ref{fig:models}.A. Each MWR-Block includes the following components: an FC layer, a layer normalization across the layers \cite{ba2016layer}, a Rectified Linear Unit (ReLU) activation function \cite{nair2010rectified}, an FC layer, a layer normalization, a ReLU activation, and finally a residual connection that sums the block's input with its final output.

The optimized base model features four MWR-Blocks, with each FC layer containing 256 units. The output layer is a single-unit FC layer with a sigmoid activation function for binary classification.

\subsection{Self-Contrastive MWR Neural Networks}
\textbf{\textit{Inward learning, not outward wandering}}: Our proposed self-contrastive models aim to optimize the embedding space by distinguishing features within individual cases, rather than relying on comparisons across multiple samples. This inward-focused approach improves predictive accuracy and generalizability.

The proposed J-MWR framework employs three interconnected self-contrastive tiers. This hierarchical approach integrates fine-grained, regional, and systemic analyses, positioning J-MWR as a potentially versatile and scalable solution for MWR breast cancer screening.

The simplest tier, Local-MWR (L-MWR), performs point-to-point comparisons across both breasts, enabling detailed intra-breast analysis to detect localized anomalies. This is particularly effective at detecting high-growth-rate tumors.

The second tier, Regional-MWR (R-MWR), extracts more complex, region-dependent features. It compares corresponding regions across the two breasts to detect symmetries or anomalies, capturing cases where the tumor is larger and/or its growth rate is slowing down.

At the highest level, the Global-MWR (G-MWR) tier utilizes features from both breasts to detect subtle asymmetries that might indicate early signs of cancer. This is achieved by swapping the values of the breasts and comparing them against the original values, relying on the fact that healthy volunteers exhibit small variations in temperature readings between the two breasts.

\subsubsection{Local-MWR Neural Network}
\label{sec:method-lmwr}
The L-MWR neural network processes each temperature point individually, excluding reference measurements, resulting in 18 inputs that consist of both skin and internal temperature values (see Figure \ref{fig:models}B). The network is structured into two main stages: feature extraction and feature comparison.

In the feature extraction stage, four 64-unit MWR-Blocks are used, followed by a single-unit FC layer with ReLU activation. The weights of the layers are shared across all inputs, ensuring consistent feature learning across the data points.

In the feature comparison stage, the network computes the absolute differences between features extracted from all pairwise inputs. These differences are filtered by setting values below a learnable threshold to zero, effectively discarding small, noisy variations. This filtering mechanism, experimentally validated to enhance performance across all proposed networks, helps focus on significant discrepancies while reducing sensitivity to noise. As we have multiple pairs, the filtered differences are then averaged to produce a mean comparison value.

The final prediction is generated using a single-unit FC layer with tanh activation, as the estimated mean value is already bounded below by zero.

\subsubsection{Regional-MWR Neural Network}
\label{sec:method-rmwr}
The R-MWR neural network performs self-contrastive analysis by comparing corresponding regions of the left and right breasts (see Figure \ref{fig:models}C). It processes two input vectors, each of size 24, which include breast-specific and reference temperature points. Similar to the L-MWR network, it consists of two main stages: feature extraction and feature comparison.

In the feature extraction stage, four MWR-Blocks are used, followed by an FC layer with ReLU activation, each containing 256 units. The weights of these layers are shared between the left and right inputs to ensure consistent feature extraction. To prevent extreme values and improve stability, the feature embeddings for each input are normalized using $l_2$-normalization.

In the feature comparison stage, element-wise absolute differences between the normalized feature embeddings of the two inputs are computed. These differences are filtered, with small values below a learnable threshold set to zero, and then summed to a single value to produce a comparison score.

The final prediction is produced by a single-unit FC layer with tanh activation.

\subsubsection{Global-MWR Neural Network}
\label{sec:method-gmwr}
The G-MWR neural network learns features from both breasts (see Figure \ref{fig:models}D). To enable self-contrastive learning, the network uses a transformed version of the original input as its second input, where the left and right breast values are swapped (i.e., the left breast values are used as the right, and vice versa). The network takes two pairs of input, each of size 44, and uses the same architecture as R-MWR, described in section \ref{sec:method-rmwr}.

\subsubsection{Joint-MWR Neural Network}
The J-MWR meta-classifier combines the pre-trained L-MWR (section \ref{sec:method-lmwr}), R-MWR (section \ref{sec:method-rmwr}), and G-MWR (section \ref{sec:method-gmwr}) models to leverage their complementary strengths (see Figure \ref{fig:models}E). Each sub-network's output is weighted using an individual FC layer and then concatenated into a unified feature vector. This combined vector is passed through a final single-unit FC layer with tanh activation to produce the final prediction. As we are only fine-tuning the weights of the sub-networks, the learning rate is reduced to \num{1e-7} to preserve the learned features while allowing gradual optimization.

\subsection{Experiments}
To ensure robustness and reliability, each model was trained and evaluated over three runs with different random initialization seeds. The results reflect the average performance across these runs on the test set.

Model evaluation was based on the Matthew's correlation coefficient (MCC), which is particularly useful for assessing performance on imbalanced datasets where both positive and negative cases are equally important \cite{chicco2020advantages}. Additionally, we report accuracy and the area under the receiver operating characteristic curve (ROC AUC) as reference metrics.

Comparative analyses were conducted between the proposed self-contrastive models and widely used batch-wise contrastive loss functions, including contrastive loss \cite{hadsell2006dimensionality}, triplet hard loss \cite{hermans2017defense}, triplet semi-hard loss \cite{schroff2015facenet}, and N-pairs loss \cite{sohn2016improved}. For these configurations, the weight for contrastive losses was experimentally set to 0.1, while cross-entropy classification loss retained a weight of 1.0.

\section{Results}
\subsection{Model Evaluation}
Our proposed J-MWR model demonstrates the highest predictive capability for identifying breast cancer from MWR data. It achieves an MCC score of 0.74 $\pm$ 0.018, outperforming the second-best model, R-MWR, by a margin of 0.08. This corresponds to an accuracy of 0.95 $\pm$ 0.003 and an ROC AUC of 0.96 $\pm$ 0.001. In comparison, the base model achieves an MCC score of 0.58 $\pm$ 0.004, an accuracy of 0.88 $\pm$ 0.003, and an ROC AUC of 0.93 $\pm$ 0.006. A summary of results for all models is presented in Table \ref{tab:results}. 

Among the individual sub-networks, both R-MWR and G-MWR outperform the base model, with MCC scores of 0.66 $\pm$ 0.012 and 0.61 $\pm$ 0.045, respectively. However, L-MWR underperforms compared to the base model, achieving an MCC score of 0.43 $\pm$ 0.002. This result aligns with expectations, as L-MWR learns features only from individual points, limiting its predictive power.

\begin{table}[!htb]
\centering
\caption{The mean and standard deviation of Matthew's correlation coefficient (MCC), accuracy, and receiver operating characteristic (ROC) area under the curve (AUC) for the base model and the proposed models: L-MWR, R-MWR, G-MWR, and J-MWR.}
\label{tab:results}
\begin{tabular}{llll}
\hline
\textbf{Model} & \textbf{MCC}         & \textbf{Accuracy}              & \textbf{ROC AUC}         \\ \hline
Base  & 0.58 $\pm$ 0.004 & 0.88 $\pm$ 0.003 &  0.93 $\pm$ 0.006 \\
L-MWR & 0.43 $\pm$ 0.002 & 0.82 $\pm$ 0.002 & 0.89 $\pm$ 0.001 \\
R-MWR & 0.66 $\pm$ 0.012 & 0.92 $\pm$ 0.003 & 0.95 $\pm$ 0.006 \\
G-MWR & 0.61 $\pm$ 0.045 & 0.90 $\pm$ 0.021 & 0.94 $\pm$ 0.016 \\
J-MWR & \textbf{0.74 $\pm$ 0.018} & \textbf{0.95 $\pm$ 0.003} & \textbf{0.96 $\pm$ 0.001}
\end{tabular}
\end{table}

\subsection{Batch-wise Contrastive Loss Evaluation}
When incorporating batch-wise contrastive loss, J-MWR with triplet hard loss achieves the highest performance among the contrastive configurations,  with an MCC score of 0.74 $\pm$ 0.03 as shown in Table \ref{tab:results_contrastive}. Overall, J-MWR consistently achieves the highest MCC score compared to other models, regardless of the batch-wise contrastive loss used. However, none of the configurations surpass J-MWR without batch-wise contrastive learning.

Interestingly, the impact of batch-wise contrastive learning varies by model. Both R-MWR and J-MWR experience a slight reduction in MCC scores when contrastive losses are applied. In contrast, L-MWR shows a modest improvement of 0.01 in MCC, while G-MWR exhibits a more substantial improvement, with MCC gains ranging from 0.03 to 0.06. Similarly, the base model benefits from applying contrastive and N-pairs losses, achieving a 0.02 increase in MCC for both configurations.

\begin{table}[!htb]
\centering
\caption{The mean and standard deviation of Matthew's correlation coefficient (MCC) results for each model trained with batch-wise contrastive losses (contrastive, N-pairs, triplet hard, and triplet semi-hard). MCC values in \textbf{bold} indicate improvements over their non-batch-wise loss counterparts.}
\label{tab:results_contrastive}
\resizebox{\columnwidth}{!}{%
\begin{tabular}{lllll}
\hline
\multicolumn{1}{c}{\multirow{2}{*}{\textbf{Model}}} & \multicolumn{4}{c}{\textbf{MCC}}                                                                                                         \\
\multicolumn{1}{c}{}                                & \multicolumn{1}{c}{Contrastive} & \multicolumn{1}{c}{N-pairs} & \multicolumn{1}{c}{Triplet hard} & \multicolumn{1}{c}{Triplet semi-hard} \\ \hline
Base                                                & \textbf{0.60 $\pm$ 0.024}                   & \textbf{0.60 $\pm$ 0.035}               & 0.57 $\pm$ 0.016                    & 0.51 $\pm$ 0.149                         \\
L-MWR                                               & \textbf{0.44 $\pm$ 0.003}                   & \textbf{0.44 $\pm$ 0.001}               & \textbf{0.44 $\pm$ 0.004}                    & \textbf{0.44 $\pm$ 0.008}                         \\
R-MWR                                               & 0.63 $\pm$ 0.060                   & 0.59 $\pm$ 0.019               & 0.60 $\pm$ 0.026                    & 0.64 $\pm$ 0.025                         \\
G-MWR                                               & \textbf{0.65 $\pm$ 0.005}                   & \textbf{0.64 $\pm$ 0.012}               & \textbf{0.64 $\pm$ 0.022}                    & \textbf{0.67 $\pm$ 0.017}                         \\
J-MWR                                               & 0.71 $\pm$ 0.000                   & 0.71 $\pm$ 0.029               & 0.74 $\pm$ 0.030                    & 0.70 $\pm$ 0.081                         
\end{tabular}%
}
\end{table}

\subsection{Embedding Space Analysis}
\begin{figure*}[!htb]
\centering
\begin{subfigure}{.4\textwidth}
  \centering
  \includegraphics[width=1.0\linewidth]{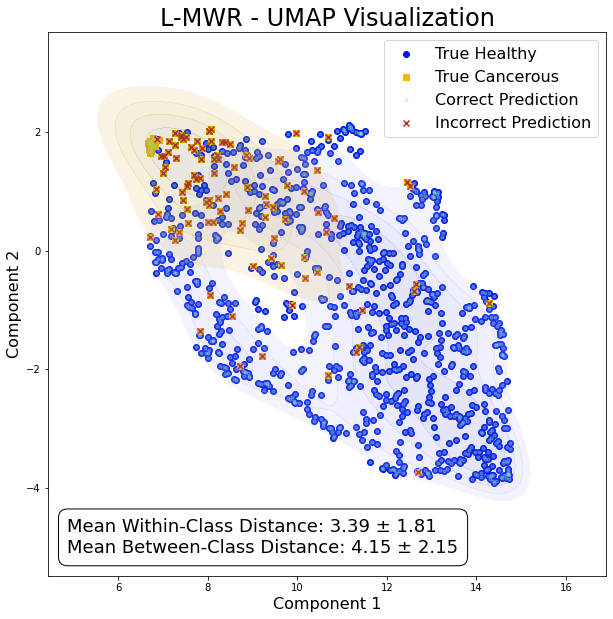}
  \caption{}
  \label{fig:umap_l_mwr}
\end{subfigure}%
\begin{subfigure}{.4\textwidth}
  \centering
  \includegraphics[width=1.0\linewidth]{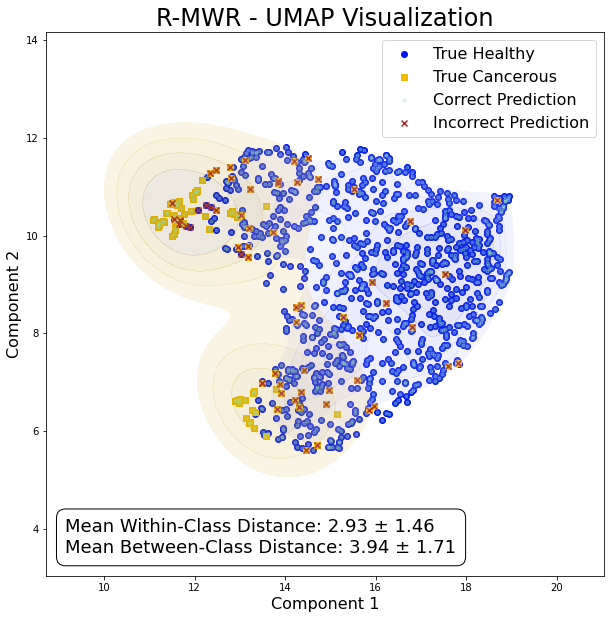}
  \caption{}
  \label{fig:umap_r_mwr}
\end{subfigure}
\begin{subfigure}{.4\textwidth}
  \centering
  \includegraphics[width=1.0\linewidth]{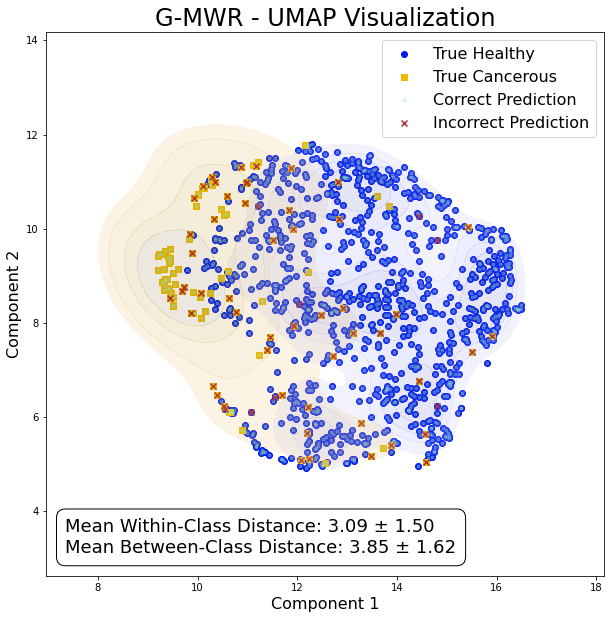}
  \caption{}
  \label{fig:umap_g_mwr}
\end{subfigure}%
\begin{subfigure}{.4\textwidth}
  \centering
  \includegraphics[width=1.0\linewidth]{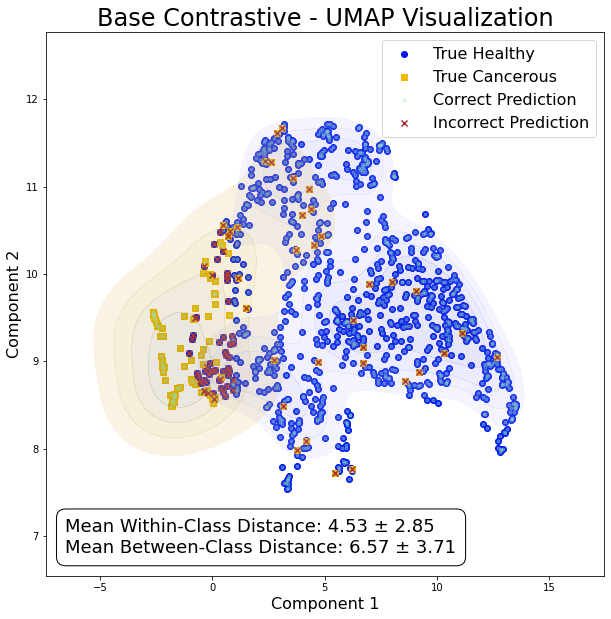}
  \caption{}
  \label{fig:umap_base_contrastive}
\end{subfigure}
\caption{Uniform Manifold Approximation and Projection (UMAP) \cite{mcinnes2018umap} visualizations of the embedding spaces generated by \textbf{(a)} L-MWR, \textbf{(b)} R-MWR, \textbf{(c)} G-MWR, and \textbf{(d)} base contrastive models. In each plot, blue circles represent correct healthy predictions, yellow squares denote correct cancerous predictions, and red crosses overlaid on these markers indicate incorrect predictions.}
\label{fig:umap}
\end{figure*}

To investigate the properties of the embedding spaces generated by the L-MWR, R-MWR, G-MWR, and base contrastive models, we used uniform manifold approximation and projection (UMAP) \cite{mcinnes2018umap} for 2D projections, as shown in Figure \ref{fig:umap}. Our analysis indicates that the base contrastive model excels at creating a clearer separation between healthy and cancerous groups, with a mean between-class Euclidean distance of $6.57 \pm 3.71$. This result is expected, as the base model is explicitly trained for this delineation. However, it exhibits greater disparities within each group, reflected in a mean within-class distance of $4.53 \pm 2.85$, contributing to its relatively lower overall performance.

In contrast, our proposed self-contrastive models demonstrate consistent embedding properties for cancerous cases across instances, even though their embeddings are more interwoven with those of healthy cases, making boundary separation less distinct. For example, R-MWR achieves a mean within-class distance of $2.93 \pm 1.46$ and a mean between-class distance of $3.94 \pm 1.17$. This closer proximity within groups mitigates the lack of a clear boundary, supporting improved classification performance. These findings suggest that self-contrastive and batch-wise contrastive approaches may be complementary. If successfully integrated together, they may further improve performance.

\subsection{Data Constraint Training}

\begin{figure*}[!htbp]
\begin{subfigure}{.5\textwidth}
  \centering
  \includegraphics[width=1.0\linewidth]{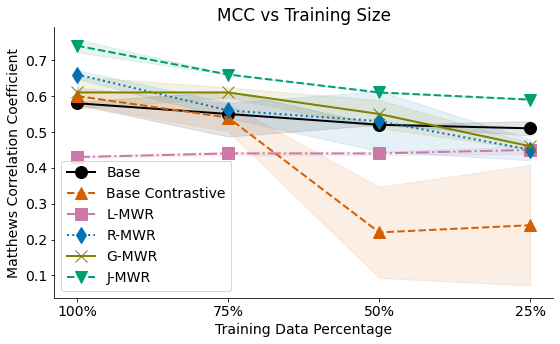}
  \caption{}
  \label{fig:data_limit_models}
\end{subfigure}%
\begin{subfigure}{.5\textwidth}
  \centering
  \includegraphics[width=1.0\linewidth]{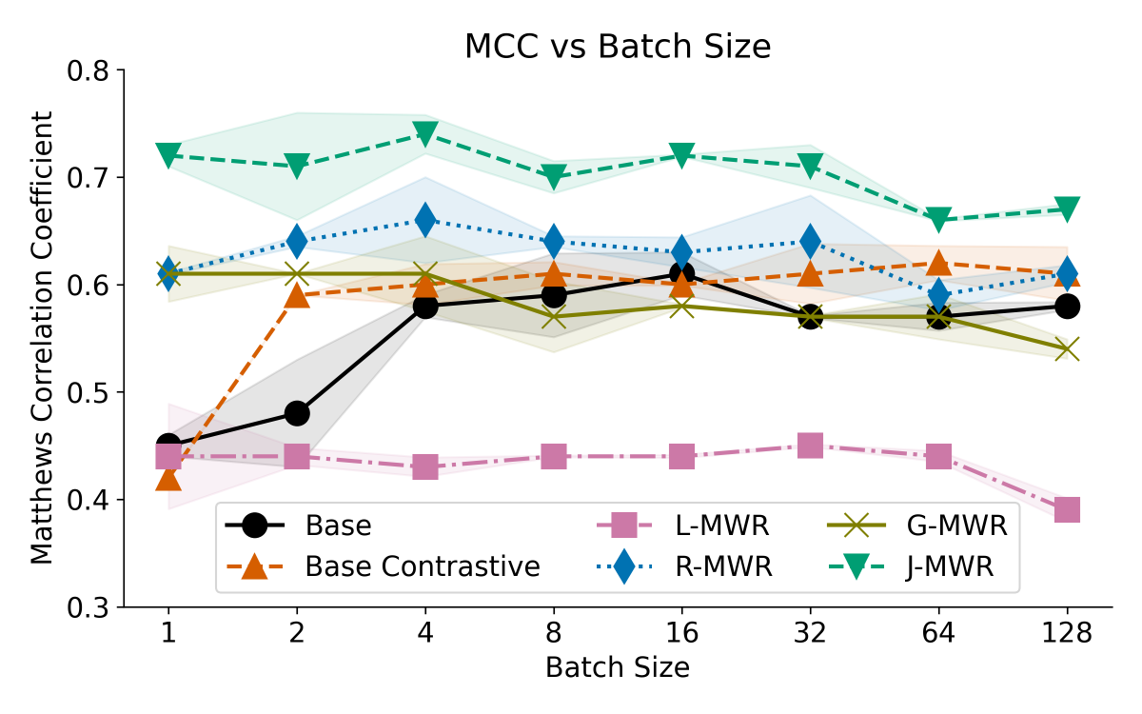}
  \caption{}
  \label{fig:batch_models}
\end{subfigure}
\caption{\textbf{(a)} Matthew's correlation coefficient (MCC) scores for each model when trained on reduced datasets with 75\%, 50\%, and 25\% of the original training size. \textbf{(b)} MCC scores for the models across varying batch sizes, ranging from 1 to 128.}
\end{figure*}

To evaluate model scalability with limited training data, we retrain the models using randomly selected subsets of the training set at 75\%, 50\%, and 25\% of the original size. This analysis provides insights into how model performance adapts with varying data availability. The results are summarized in Figure \ref{fig:data_limit_models}. J-MWR consistently achieves the highest MCC across all subsets, with its lowest performance recorded at $0.69 \pm 0.001$ at 25\% of the data. In contrast, the base contrastive model shows the weakest performance when 50\% or less of the training data is used, but noticeably improves at 75\%.

Interestingly, L-MWR maintains a stable but low performance level regardless of the reduction in training data size, reflecting its limited reliance on larger datasets. R-MWR, on the other hand, achieves the highest performance improvement, with a 0.1 MCC increase when training data is expanded from 75\% to 100\%. This is also reflected on J-MWR, which depends on R-MWR, though the impact is less pronounced. If this trend continues with even larger datasets, R-MWR has the potential to surpass the combined performance of J-MWR.

For the other models, performance tends to plateau as more data is added, indicating limited additional benefits from larger datasets for these configurations.


%


\subsection{Batch Size Dependency}
\label{sec:results_batch}
We analyze the effect of batch size on performance across values ranging from 1 to 128. J-MWR consistently outperforms all other models, achieving its highest MCC score of $0.74 \pm 0.018$ at a batch size of 4. Beyond this point, performance gradually declines, as illustrated in Figure \ref{fig:batch_models}. Overall, the proposed self-contrastive models show reduced performance as the batch size increases.

In contrast, the base contrastive model, followed by the base model, maintains more consistent performance after a batch size of 4. We anticipate this trend will extend to larger batch sizes. This stability highlights a significant distinction in the behavior of batch-wise contrastive losses compared to self-contrastive models.

For self-contrastive models, smaller batch sizes are optimal for accuracy but introduce a trade-off with longer training times. Meanwhile, batch-wise contrastive approaches for MWR show minimal dependency on batch size, making them more efficient for scenarios requiring larger batches or faster training.

\subsection{Generalizability}

\begin{figure*}[ht!]
\begin{subfigure}{.5\textwidth}
  \centering
  \includegraphics[width=1.0\linewidth]{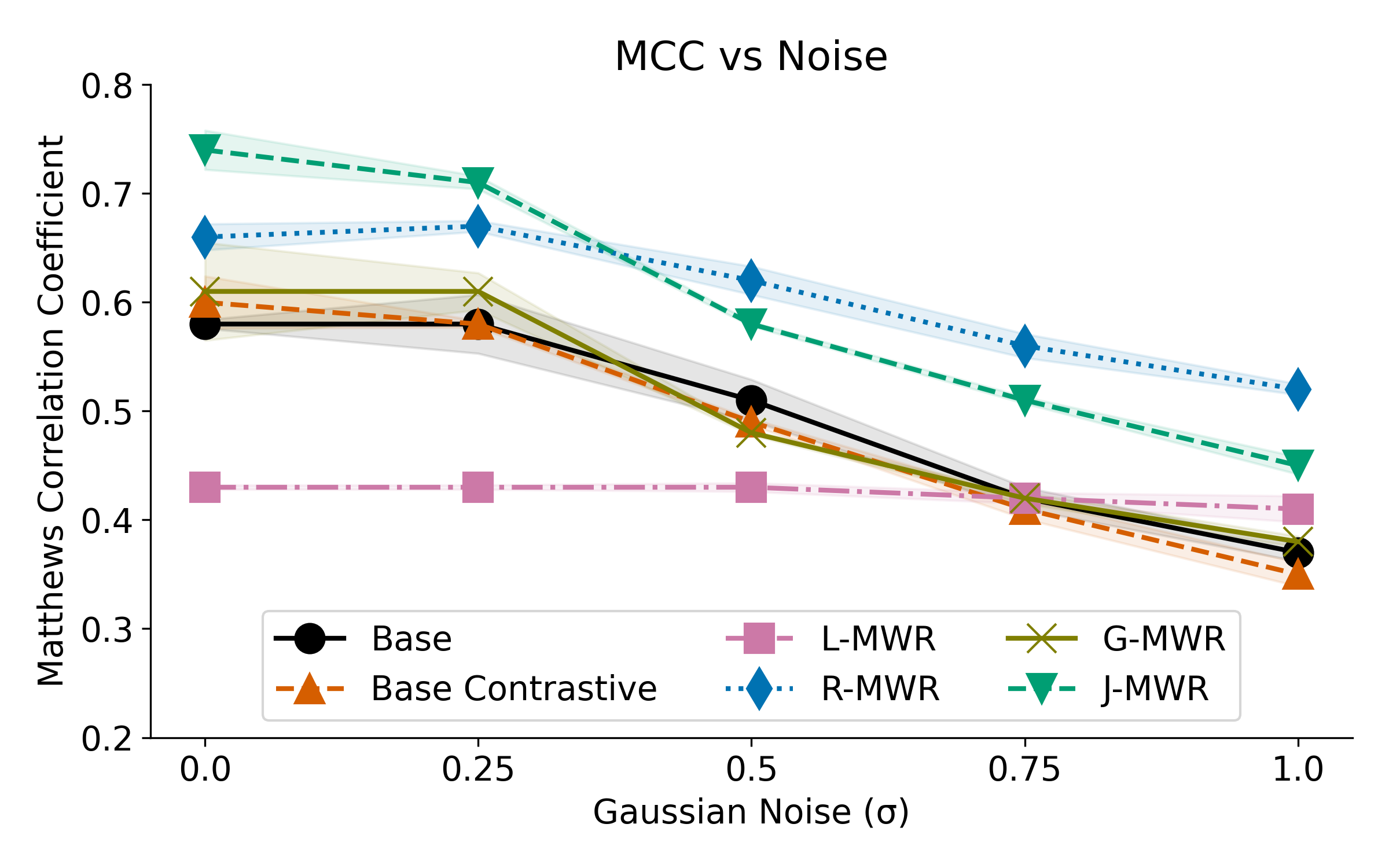}
  \caption{}
  \label{fig:data_noise}
\end{subfigure}%
\begin{subfigure}{.5\textwidth}
  \centering
  \includegraphics[width=1.0\linewidth]{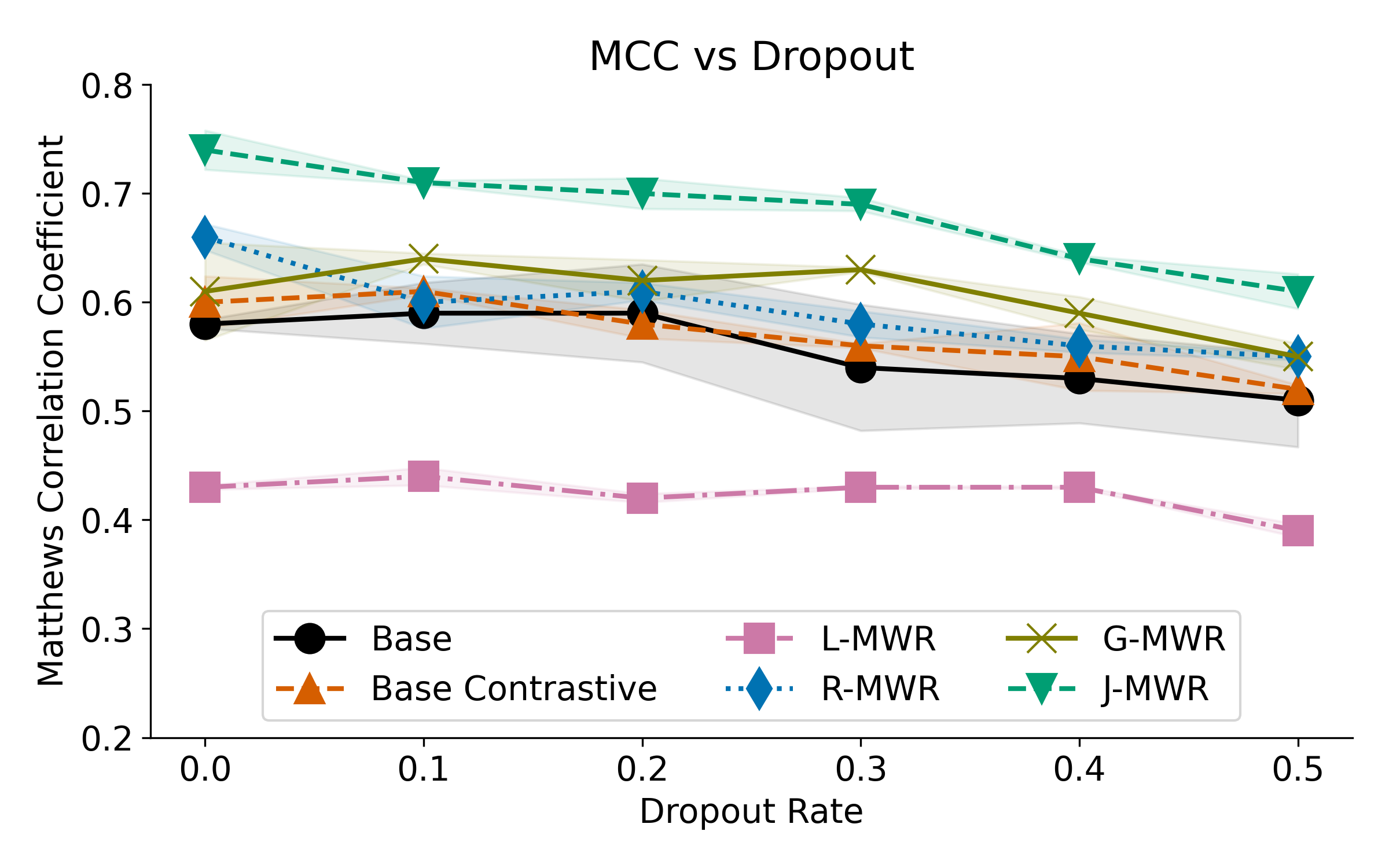}
  \caption{}
  \label{fig:data_dropout}
\end{subfigure}
\begin{subfigure}{.5\textwidth}
  \centering
  \includegraphics[width=1.0\linewidth]{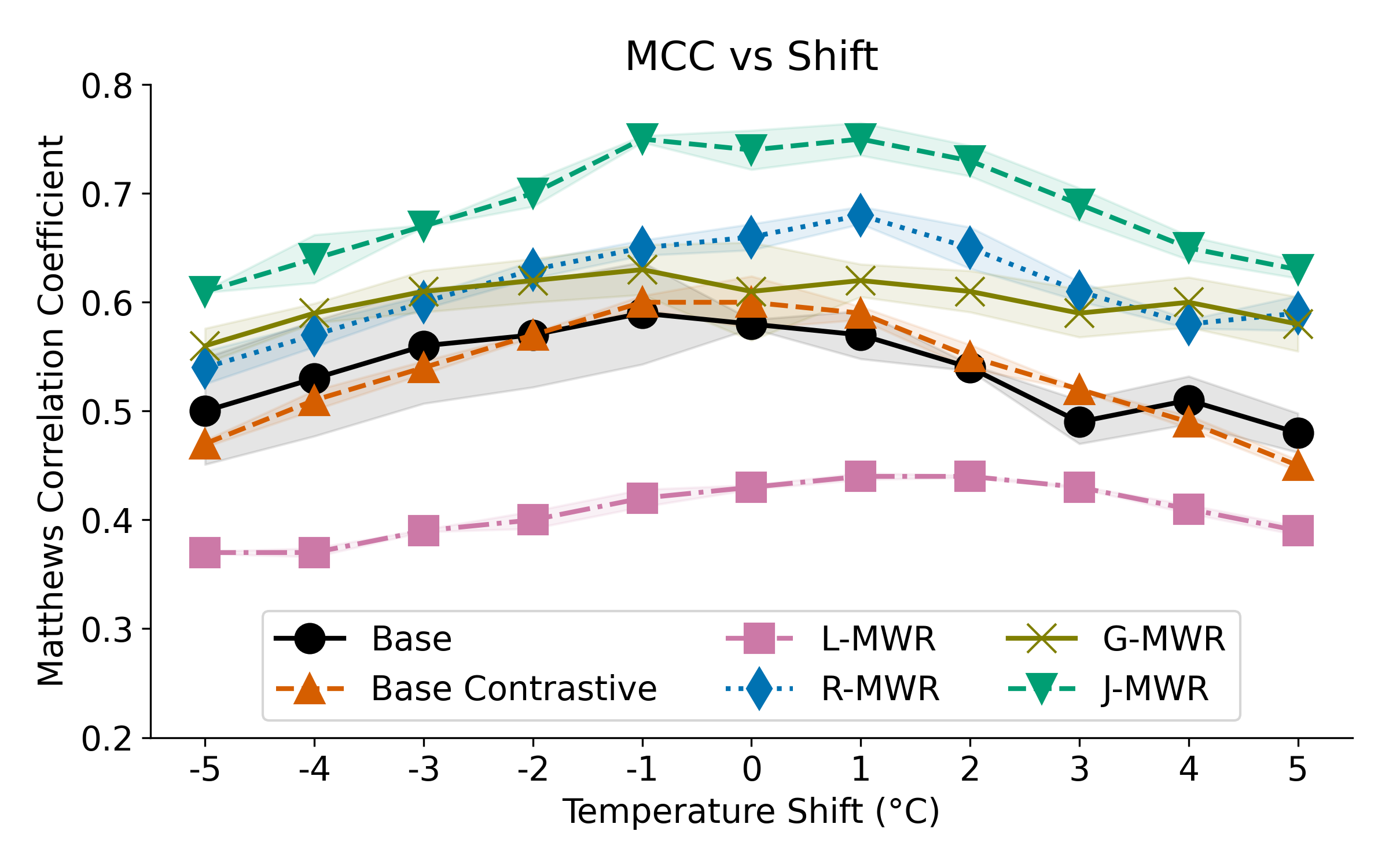}
  \caption{}
  \label{fig:data_shift}
\end{subfigure}%
\begin{subfigure}{.5\textwidth}
  \centering
  \includegraphics[width=1.0\linewidth]{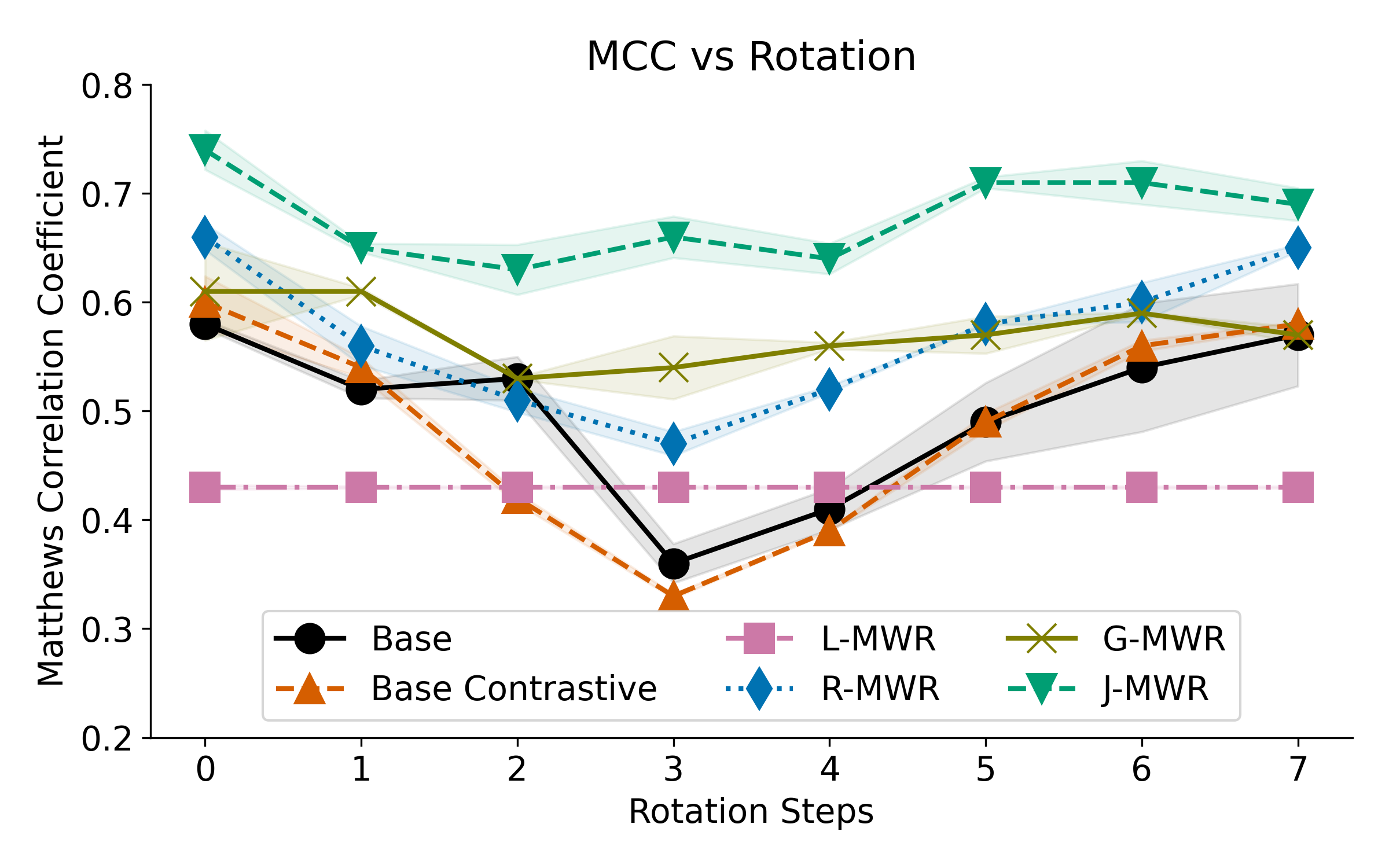}
  \caption{}
  \label{fig:data_rotation}
\end{subfigure}
\caption{Matthew's correlation coefficient (MCC) scores of the models under various data augmentations: \textbf{(a)} adding increasing levels of Gaussian noise, \textbf{(b)} randomly setting points to the mean of the remaining points, \textbf{(c)} uniformly shifting all temperatures by a given amount, and \textbf{(d)} rotating the points around the nipple for each breast.}
\label{fig:data_augmentation}
\end{figure*}

To evaluate the generalizability of our trained models, we tested them on an augmented dataset featuring a range of out-of-distribution transformations. These include the addition of Gaussian noise (Figure \ref{fig:data_noise}), applying dropout to temperature points by replacing their values with the mean of the remaining points (Figure \ref{fig:data_dropout}), global temperature shifts across all values (Figure \ref{fig:data_shift}), and rotations of breast points around the nipple (Figure \ref{fig:data_rotation}).

As shown in Figure \ref{fig:data_augmentation}, J-MWR consistently demonstrates better generalization performance, maintaining a higher MCC score across various transformations. This holds true for both data corruptions (Figures \ref{fig:data_noise} and \ref{fig:data_dropout}) and data drifts (Figures \ref{fig:data_shift} and \ref{fig:data_rotation}). However, under severe Gaussian noise ($\sigma > 0.25$), R-MWR exhibits greater resilience, as J-MWR's performance is limited by the degraded performance of G-MWR under noisy conditions.

While L-MWR shows the least variation across transformations, its MCC remains low at approximately 0.43. Interestingly, under certain conditions (see Figure \ref{fig:data_augmentation}), it surpasses the base, base contrastive, and/or G-MWR models.

Overall, these results highlight the robustness of J-MWR in managing diverse data augmentations, with a degree of sensitivity to specific noise types. Its ability to generalize to unknown distributions stems from its focus on comparing regions within individual cases, rather than relying on comparisons across samples or the training distribution.

\subsection{Ensemble Methods}
\label{sec:results_ensemble}

\begin{figure*}[thb!]
\centering
  \includegraphics[width=0.5\textwidth]{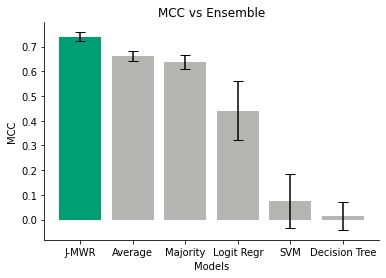}
  \caption{Matthew's correlation coefficient (MCC) scores for ensemble methods, including J-MWR, averaging, majority voting, logistic regression, support vector machine (SVM), and decision tree, leveraging the pre-trained L-MWR, R-MWR, and G-MWR models.}
  \label{fig:ensemble}
\end{figure*}

Each sub-network in the J-MWR meta-classifier contributes nearly equally to the final prediction.
The L-MWR has a contribution weight of 0.998, comparable to R-MWR and G-MWR, which both have weights much closer to 1.0. Additionally, the weight biases of the sub-networks are close to zero. As illustrated in Figure \ref{fig:ensemble}, J-MWR consistently outperforms alternative ensemble techniques, including averaging and majority voting, as well as meta-classifiers such as logistic regression, SVM, and decision tree. 

While J-MWR bears a resemblance to average voting, given the previous weights, the fine-tuning applied to its sub-networks significantly boosts MCC performance. Averaging the predictions of L-MWR, R-MWR, and G-MWR achieves the second-best MCC score of $0.66 \pm 0.02$, which aligns closely with the performance of R-MWR alone. Majority voting follows closely behind in performance.

In contrast, meta-classifiers such as logistic regression, SVM, and decision tree demonstrate overfitting during training, leading to large drops in MCC performance on the test set. These results emphasize the effectiveness of J-MWR in leveraging its fine-tuned sub-networks to deliver improved predictive capabilities compared to traditional ensemble methods.

\section{Discussion}
This study successfully demonstrates how self-contrastive learning can improve the performance of MWR breast cancer detection. The proposed architectures, L-MWR, R-MWR, and G-MWR, integrate hand-engineered features capturing thermal asymmetries with data-driven learning. Our meta-classifier, J-MWR, combines the results of the individual models and achieves an MCC score of $0.74 \pm 0.018$, outperforming all existing models. This highlights the efficacy of self-contrastive learning in enhancing both accuracy and generalizability for breast cancer detection, making it a promising advancement in the field.

The final layer of J-MWR effectively acts as an ensemble averaging mechanism, given the near-equal weights of each sub-network. However, J-MWR achieves an MCC improvement of 0.08 compared to traditional ensemble averaging. This boost stems from its role as a fine-tuning meta-classifier, enabling indirect information sharing across the self-contrastive tiers. While this design architecture boosts performance, it comes with the trade-off of increased model complexity and training time.

Furthermore, batch-wise contrastive learning, while yielding modest improvements, holds potential as a minimally disruptive addition to MWR models. Its current subpar performance can be attributed to the inherent variability in breast temperature readings, influenced by factors such as age and menstrual cycle phase \cite{goryanin2020passive}. By addressing these nuances, batch-wise contrastive learning could further complement self-contrastive approaches, contributing to the development of more robust diagnostic models.

While self-contrastive learning shows promising results for MWR breast analysis, it has certain limitations. It is less effective for patients with mastectomies, reconstructive surgeries, or hormonal therapies, which alter thermal profiles and reduce the reliability of asymmetry-based comparisons. Its extension to anatomical regions lacking symmetrical properties is also challenging, limiting its broader applicability. Moreover, the approach demands manual definition of sub-networks, which is time-intensive and requires greater domain expertise compared to typical supervised learning. Additionally, the increased model complexity may pose challenges in deployment on resource-constrained devices, potentially restricting its use in low-resource settings. Addressing these limitations will be crucial for improving the versatility and accessibility of MWR self-contrastive learning frameworks, especially in point-of-care settings.

To address these limitations, future work will explore the integration of representation learning methods \cite{assran2023self} and the use of neural architecture search to automate sub-network design \cite{li2022dynamic}, reducing reliance on domain expertise. Additionally, sampling strategies tailored to physiological factors such as age and hormonal cycles will be developed to refine positive and negative pair selections, further improving the robustness and generalizability of the framework. Meta-data sampling strategies have been successfully applied in other medical domains \cite{holland2024metadata,gorade2023pacl,vu2021medaug}. Additionally, we will integrate multi-modal data, such as patient health records, miRNA, and other multi-omics data \cite{fisher2022passive}. Finally, it is crucial we evaluate our proposed model across various anatomical locations and under different physiological conditions to assess its broad applicability and effectiveness in diverse clinical settings.

\section{Conclusion}
Our research introduces promising advancements in MWR breast cancer detection, highlighting significant potential for clinical applications and future exploration. The proposed models, particularly the J-MWR architecture, demonstrate strong performance in accurately identifying breast cancer with high generalizability and robustness. By combining thermal asymmetry analysis with data-driven methodologies, we bridge the gap between traditional diagnostic approaches and modern machine learning techniques, paving the way for more efficient and accurate breast cancer detection. The integration of multi-modal data could further improve model personalization of the system. While challenges remain, particularly in ensuring the models’ applicability to all patient demographics, future work will focus on overcoming these barriers, advancing the technology for wider, more equitable use.


\bibliographystyle{IEEEtran}
\bibliography{main}

\begin{thebibliography}{10}
\providecommand{\url}[1]{#1}
\csname url@samestyle\endcsname
\providecommand{\newblock}{\relax}
\providecommand{\bibinfo}[2]{#2}
\providecommand{\BIBentrySTDinterwordspacing}{\spaceskip=0pt\relax}
\providecommand{\BIBentryALTinterwordstretchfactor}{4}
\providecommand{\BIBentryALTinterwordspacing}{\spaceskip=\fontdimen2\font plus
\BIBentryALTinterwordstretchfactor\fontdimen3\font minus \fontdimen4\font\relax}
\providecommand{\BIBforeignlanguage}[2]{{%
\expandafter\ifx\csname l@#1\endcsname\relax
\typeout{** WARNING: IEEEtran.bst: No hyphenation pattern has been}%
\typeout{** loaded for the language `#1'. Using the pattern for}%
\typeout{** the default language instead.}%
\else
\language=\csname l@#1\endcsname
\fi
#2}}
\providecommand{\BIBdecl}{\relax}
\BIBdecl

\bibitem{sung2021global}
H.~Sung, J.~Ferlay, R.~L. Siegel, M.~Laversanne, I.~Soerjomataram, A.~Jemal, and F.~Bray, ``Global cancer statistics 2020: Globocan estimates of incidence and mortality worldwide for 36 cancers in 185 countries,'' \emph{CA: a cancer journal for clinicians}, vol.~71, no.~3, pp. 209--249, 2021.

\bibitem{arnold2022current}
M.~Arnold, E.~Morgan, H.~Rumgay, A.~Mafra, D.~Singh, M.~Laversanne, J.~Vignat, J.~R. Gralow, F.~Cardoso, S.~Siesling, and I.~Soerjomataram, ``Current and future burden of breast cancer: Global statistics for 2020 and 2040,'' \emph{The Breast}, vol.~66, pp. 15--23, 2022.

\bibitem{goryanin2020passive}
I.~Goryanin, S.~Karbainov, O.~Shevelev, A.~Tarakanov, K.~Redpath, S.~Vesnin, and Y.~Ivanov, ``Passive microwave radiometry in biomedical studies,'' \emph{Drug Discovery Today}, vol.~25, no.~4, pp. 757--763, 2020.

\bibitem{vesnin2017modern}
S.~Vesnin, A.~K. Turnbull, J.~M. Dixon, and I.~Goryanin, ``Modern microwave thermometry for breast cancer,'' \emph{J. Mol. Imaging Dyn}, vol.~7, no.~2, p. 1000136, 2017.

\bibitem{fisher2022passive}
L.~Fisher, O.~Fisher, D.~Chebanov, S.~Vesnin, A.~Goltsov, A.~Turnbull, M.~Dixon, I.~Kudaibergenova, B.~Osmonov, S.~Karbainov, L.~Popov, A.~Losev, and I.~Goryanin, ``Passive microwave radiometry and microrna detection for breast cancer diagnostics,'' \emph{Diagnostics}, vol.~13, no.~1, p. 118, 2022.

\bibitem{khoperskov2022improving}
A.~V. Khoperskov and M.~V. Polyakov, ``Improving the efficiency of oncological diagnosis of the breast based on the combined use of simulation modeling and artificial intelligence algorithms,'' \emph{Algorithms}, vol.~15, no.~8, p. 292, 2022.

\bibitem{wang2023microwave}
L.~Wang, ``Microwave imaging and sensing techniques for breast cancer detection,'' \emph{Micromachines}, vol.~14, no.~7, p. 1462, 2023.

\bibitem{bhargava2024microwave}
D.~Bhargava, P.~Rattanadecho, and K.~Jiamjiroch, ``Microwave imaging for breast cancer detection-a comprehensive review,'' \emph{Engineered Science}, vol.~30, p. 1116, 2024.

\bibitem{shevelev2022using}
O.~Shevelev, M.~Petrova, A.~Smolensky, B.~Osmonov, S.~Toimatov, T.~Kharybina, S.~Karbainov, L.~Ovchinnikov, S.~Vesnin, A.~Tarakanov, and I.~Goryanin, ``Using medical microwave radiometry for brain temperature measurements,'' \emph{Drug discovery today}, vol.~27, no.~3, pp. 881--889, 2022.

\bibitem{shevelev2023correction}
O.~A. Shevelev, M.~V. Petrova, E.~M. Mengistu, M.~Y. Yuriev, I.~Z. Kostenkova, S.~G. Vesnin, M.~M. Kanarskii, M.~A. Zhdanova, and I.~Goryanin, ``Correction of local brain temperature after severe brain injury using hypothermia and medical microwave radiometry (mwr) as companion diagnostics,'' \emph{Diagnostics}, vol.~13, no.~6, p. 1159, 2023.

\bibitem{shevelev2023diagnostics}
O.~A. Shevelev, A.~V. Smolensky, M.~V. Petrova, E.~M. Mengistu, A.~A. Mengistu, M.~V. VatsikGorodetskaya, U.~G. Khanakhmedova, D.~N. Menzhurenkova, S.~G. Vesnin, and I.~I. Goryanin, ``Diagnostics and prevention of sports-related traumatic brain injury complication,'' \emph{RUDN Journal of Medicine}, vol.~27, no.~2, pp. 254--264, 2023.

\bibitem{hossain2023lightweight}
A.~Hossain, M.~T. Islam, S.~K. Abdul~Rahim, M.~A. Rahman, T.~Rahman, H.~Arshad, A.~Khandakar, M.~A. Ayari, and M.~E. Chowdhury, ``A lightweight deep learning based microwave brain image network model for brain tumor classification using reconstructed microwave brain (rmb) images,'' \emph{Biosensors}, vol.~13, no.~2, p. 238, 2023.

\bibitem{osmonov2021passive}
B.~Osmonov, L.~Ovchinnikov, C.~Galazis, B.~Emilov, M.~Karaibragimov, M.~Seitov, S.~Vesnin, A.~Losev, V.~Levshinskii, I.~Popov, C.~Mustafin, T.~Kasymbekov, and I.~Goryanin, ``Passive microwave radiometry for the diagnosis of coronavirus disease 2019 lung complications in kyrgyzstan,'' \emph{Diagnostics}, vol.~11, no.~2, p. 259, 2021.

\bibitem{emilov2023diagnostic}
B.~Emilov, A.~Sorokin, M.~Seiitov, B.~T. Kobayashi, T.~Chubakov, S.~Vesnin, I.~Popov, A.~Krylova, and I.~Goryanin, ``Diagnostic of patients with covid-19 pneumonia using passive medical microwave radiometry (mwr),'' \emph{Diagnostics}, vol.~13, no.~15, p. 2585, 2023.

\bibitem{levshinskii2022using}
V.~Levshinskii, C.~Galazis, A.~Losev, T.~Zamechnik, T.~Kharybina, S.~Vesnin, and I.~Goryanin, ``Using ai and passive medical radiometry for diagnostics (mwr) of venous diseases,'' \emph{Computer Methods and Programs in Biomedicine}, vol. 215, p. 106611, 2022.

\bibitem{tarakanov2022passive}
A.~V. Tarakanov, E.~S. Ladanova, A.~A. Lebedenko, T.~D. Tarakanova, S.~G. Vesnin, T.~Kharybina, and I.~I. Goryanin, ``Passive microwave radiometry as a component of imaging diagnostics in juvenile idiopathic arthritis,'' \emph{Rheumato}, vol.~2, no.~3, pp. 55--68, 2022.

\bibitem{tarakanov2023age}
A.~V. Tarakanov, A.~A. Tarakanov, E.~G. Skorodumova, N.~Roberts, T.~Kobayshi, S.~G. Vesnin, V.~Zelman, and I.~Goryanin, ``Age-related changes in the temperature of the lumbar spine measured by passive microwave radiometry (mwr),'' \emph{Diagnostics}, vol.~13, no.~21, p. 3294, 2023.

\bibitem{levshinskii2020application}
V.~Levshinskii, C.~Galazis, L.~Ovchinnikov, S.~Vesnin, A.~Losev, and I.~Goryanin, ``Application of data mining and machine learning in microwave radiometry (mwr),'' in \emph{Biomedical Engineering Systems and Technologies: 12th International Joint Conference, BIOSTEC 2019, Prague, Czech Republic, February 22--24, 2019, Revised Selected Papers 12}.\hskip 1em plus 0.5em minus 0.4em\relax Springer, 2020, pp. 265--288.

\bibitem{galazis2019application}
C.~Galazis, S.~Vesnin, and I.~Goryanin, ``Application of artificial intelligence in microwave radiometry (mwr).'' in \emph{Bioinformatics}, 2019, pp. 112--122.

\bibitem{losev2022some}
A.~Losev, I.~Popov, A.~Y. Petrenko, A.~Gudkov, S.~Vesnin, and S.~Chizhikov, ``Some methods for substantiating diagnostic decisions made using machine learning algorithms,'' \emph{Biomedical Engineering}, vol.~55, no.~6, p. 442, 2022.

\bibitem{losev2021neural}
A.~G. Losev, D.~A. Medevedev, and A.~V. Svetlov, ``Neural networks in diagnosis of breast cancer,'' in \emph{"Smart Technologies" for Society, State and Economy 13}.\hskip 1em plus 0.5em minus 0.4em\relax Springer, 2021, pp. 220--227.

\bibitem{losev2022artificial}
A.~G. Losev and A.~V. Svetlov, ``Artificial intelligence algorithms in diagnosis of breast cancer,'' in \emph{New Technology for Inclusive and Sustainable Growth: Perception, Challenges and Opportunities}.\hskip 1em plus 0.5em minus 0.4em\relax Springer, 2022, pp. 175--182.

\bibitem{li2022dynamic}
J.~Li, C.~Galazis, L.~Popov, L.~Ovchinnikov, T.~Kharybina, S.~Vesnin, A.~Losev, and I.~Goryanin, ``Dynamic weight agnostic neural networks and medical microwave radiometry (mwr) for breast cancer diagnostics,'' \emph{Diagnostics}, vol.~12, no.~9, p. 2037, 2022.

\bibitem{germashev2021fuzzy}
I.~Germashev, V.~Dubovskaya, A.~Losev, and I.~Popov, ``Fuzzy inference of the effectiveness factors of the computational model for the diagnosis of breast cancer,'' in \emph{2021 3rd International Conference on Control Systems, Mathematical Modeling, Automation and Energy Efficiency (SUMMA)}.\hskip 1em plus 0.5em minus 0.4em\relax IEEE, 2021, pp. 528--533.

\bibitem{germashev2023hierarchical}
I.~Germashev, V.~Dubovskaya, and A.~Losev, ``Hierarchical fuzzy inference of adequacy of highly informative diagnostic signs of breast cancer,'' in \emph{Society 5.0: Cyber-Solutions for Human-Centric Technologies}.\hskip 1em plus 0.5em minus 0.4em\relax Springer, 2023, pp. 31--41.

\bibitem{zenovich2016algoritmy}
A.~Zenovich, V.~Glazunov, A.~Oparin, and F.~Primachenko, ``Algoritmy prinyatiya resheniy v konsultativnoy intellektualnoy sisteme diagnostiki molochnykh zhelez [algorithms of decision-making in intelligent advisory system for diagnostics of the mammary glands],'' \emph{Vestnik Volgogradskogo gosudarstvennogo universiteta. Seriya 1, Matematika. Fizika}, pp. 129--142, 2016.

\bibitem{losev2017intellektualnyy}
A.~Losev and V.~Levshinskiy, ``Intellektualnyy analiz termometricheskikh dannykh v diagnostike molochnykh zhelez [the thermometry data mining in the diagnostics of mammary glands],'' \emph{Upravlenie bolshimi sistemami}, no.~70, pp. 113--135, 2017.

\bibitem{glorot2010understanding}
X.~Glorot and Y.~Bengio, ``Understanding the difficulty of training deep feedforward neural networks,'' in \emph{Proceedings of the thirteenth international conference on artificial intelligence and statistics}.\hskip 1em plus 0.5em minus 0.4em\relax JMLR Workshop and Conference Proceedings, 2010, pp. 249--256.

\bibitem{kingma2014adam}
D.~P. Kingma and J.~Ba, ``Adam: A method for stochastic optimization,'' \emph{arXiv preprint arXiv:1412.6980}, 2014.

\bibitem{ba2016layer}
J.~L. Ba, ``Layer normalization,'' \emph{arXiv preprint arXiv:1607.06450}, 2016.

\bibitem{nair2010rectified}
V.~Nair and G.~E. Hinton, ``Rectified linear units improve restricted boltzmann machines,'' in \emph{Proceedings of the 27th international conference on machine learning (ICML-10)}, 2010, pp. 807--814.

\bibitem{chicco2020advantages}
D.~Chicco and G.~Jurman, ``The advantages of the matthews correlation coefficient (mcc) over f1 score and accuracy in binary classification evaluation,'' \emph{BMC genomics}, vol.~21, pp. 1--13, 2020.

\bibitem{hadsell2006dimensionality}
R.~Hadsell, S.~Chopra, and Y.~LeCun, ``Dimensionality reduction by learning an invariant mapping,'' in \emph{2006 IEEE computer society conference on computer vision and pattern recognition (CVPR'06)}, vol.~2.\hskip 1em plus 0.5em minus 0.4em\relax IEEE, 2006, pp. 1735--1742.

\bibitem{hermans2017defense}
A.~Hermans, L.~Beyer, and B.~Leibe, ``In defense of the triplet loss for person re-identification,'' \emph{arXiv preprint arXiv:1703.07737}, 2017.

\bibitem{schroff2015facenet}
F.~Schroff, D.~Kalenichenko, and J.~Philbin, ``Facenet: A unified embedding for face recognition and clustering,'' in \emph{Proceedings of the IEEE conference on computer vision and pattern recognition}, 2015, pp. 815--823.

\bibitem{sohn2016improved}
K.~Sohn, ``Improved deep metric learning with multi-class n-pair loss objective,'' \emph{Advances in neural information processing systems}, vol.~29, 2016.

\bibitem{mcinnes2018umap}
L.~McInnes, J.~Healy, and J.~Melville, ``Umap: Uniform manifold approximation and projection for dimension reduction,'' \emph{arXiv preprint arXiv:1802.03426}, 2018.

\bibitem{assran2023self}
M.~Assran, Q.~Duval, I.~Misra, P.~Bojanowski, P.~Vincent, M.~Rabbat, Y.~LeCun, and N.~Ballas, ``Self-supervised learning from images with a joint-embedding predictive architecture,'' in \emph{Proceedings of the IEEE/CVF Conference on Computer Vision and Pattern Recognition}, 2023, pp. 15\,619--15\,629.

\bibitem{holland2024metadata}
R.~Holland, O.~Leingang, H.~Bogunovi{\'c}, S.~Riedl, L.~Fritsche, T.~Prevost, H.~P. Scholl, U.~Schmidt-Erfurth, S.~Sivaprasad, A.~J. Lotery \emph{et~al.}, ``Metadata-enhanced contrastive learning from retinal optical coherence tomography images,'' \emph{Medical Image Analysis}, vol.~97, p. 103296, 2024.

\bibitem{gorade2023pacl}
V.~Gorade, S.~Mittal, and R.~Singhal, ``Pacl: Patient-aware contrastive learning through metadata refinement for generalized early disease diagnosis,'' \emph{Computers in Biology and Medicine}, vol. 167, p. 107569, 2023.

\bibitem{vu2021medaug}
Y.~N.~T. Vu, R.~Wang, N.~Balachandar, C.~Liu, A.~Y. Ng, and P.~Rajpurkar, ``Medaug: Contrastive learning leveraging patient metadata improves representations for chest x-ray interpretation,'' in \emph{Machine Learning for Healthcare Conference}.\hskip 1em plus 0.5em minus 0.4em\relax PMLR, 2021, pp. 755--769.

\end{thebibliography}

\end{document}